\documentclass[aps,prl,twocolumn,groupedaddress,floatfix,showpacs,showkeys,amsmath,amssymb]{revtex4-1}
\usepackage{graphicx}
\usepackage{bm}
\begin{document}
\title{Effect of atomic-scale defects and dopants on phosphorene electronic structure and quantum transport properties}

\author{Alejandro Lopez-Bezanilla$^{1}$}
\email[]{alejandrolb@gmail.com}
\affiliation{$^{1}$Argonne National Laboratory, 9700 S. Cass Avenue, Lemont, Illinois, 60439, United States}

\begin{abstract}
By means of a multi-scale first-principles approach, a description of the local electronic structure of two-dimensional and narrow phosphorene sheets with various types of modifications is presented. 
Firstly, a rational argument based on the geometry of the pristine and modified P network, and supported by the Wannier functions formalism is introduced to describe a hybridization model of the P atomic orbitals.
{\it Ab initio} calculations show that non-isoelectronic foreign atoms form quasi-bound states at varying energy levels and create different polarization states depending on the number of valence electrons between P and the doping atom. 
The quantum transport properties of modified phosphorene ribbons are further described with great accuracy.
The distortions on the electronic bands induced by the external species lead to strong backscattering effects on the propagating charge carriers. 
Depending on the energy of the charge-carrier and the type of doping, the conduction may range from the diffusive to the localized regime. Interstitial defects at vacant sites lead to homogeneous transport fingerprints across different types of doping atoms. We suggest that the relatively low values of charge mobility reported in experimental measurements may have their origin in the presence of defects.
\end{abstract}
\maketitle

\section{Introduction}
Phosphorene is a single layer of black phosphorus where P atoms arranged in a hexagonal staggered lattice can be exfoliated down to ultrathin layers.
In a like manner to other single-layered two-dimensional (2D) materials, phosphorene exhibits radically different electronic, mechanical, and thermal properties standing alone compared to when it is stuck with other layers forming a bulk. Moreover phosphorene has several advantages over competing thin-layered materials in the race to integrate the next generation of electronic devices \cite{doi:10.1021/nn501226z}. 

The direct electronic band gap of 2 eV \cite{PhysRevB.89.235319}, and an improved transistor performance as compared with graphene allows phosphorene to surpass all-C based transistors in terms of on/off current ratio and current saturation \cite{ISI:000340623400003}.
Due to phosphorene's ability to conduct electricity with both electrons and holes (ambipolar), it can be employed to construct p-n junctions. Charge mobility is temperature- and thickness- dependent, with experimentally measured charge-carrier mobilities peaking around 4000 cm$^2$V$^{-1}$s$^{-1}$ on 10nm thick samples \cite{2053-1583-2-1-011001} and on/off current ratios of up to $\sim$10$^5$. Measurements of the transfer characteristics of phosphorene field-effect transistors by Das et al. demonstrated that the transport gap of phosphorene can be scaled up from $\sim$0.3 to $\sim$1.0 eV when the material thickness is scaled down from bulk to a single layer \cite{doi:10.1021/nl5025535}. 

However, the pervasive presence of defects, ranging from vacancies to external atoms or functional groups that in substitution or attached to the material surface modify the intrinsic properties of the pristine material, inevitably alter the performance of the transistor. The presence of scattering centers hinders the propagation of charge carriers and reduces mobility at several orders of magnitude. But defects may also boost the integration of phosphorene in field-effect transistor devices playing an important role in the design and practical application of 2D materials. Indeed, the high surface-volume ratio of monolayered materials allows for their utilization in, for instance, chemical sensing devices. Transport measurements have proven the superior capability of phosphorene in the detection of NO$_2$ molecules based on the charge transfer from the P atoms to the external molecules \cite{ISI:000355383000099}.

In this paper we report the effect of doping, external atom attachment, and structural defects in the electronic structure of 2D and 1D phosphorene monolayers and the subsequent impact on the transport properties of micrometer-long ribbons. First-principles calculations show that non-isoelectronic foreign atoms create quasi-bound states at varying energy levels and with different polarization, depending on the difference in the number of valence electrons between P and the doping atom. Although the substitution or the attachment of external species induce small distortions of the atomic network and of the electronic bands, strong backscattering effects on the propagating charge carriers can be observed. As a result of the interaction between the charge carriers and the impurities the conduction may range from ballistic to diffusive or localized regimes, creating transport gaps that effectively enlarge the intrinsic electronic band gap of phosphorene. Our findings are in line with recent experimental results reporting lower mobilities and the charge-carrier transmission ability of black phosphorus sheets with respect to predicted values  \cite{ISI:000340623400003}.

 \section{Methodology}
 
\subsection{ {\it Ab initio} scheme}
The geometry optimizations and electronic structure calculations were carried out using the SIESTA density functional theory (DFT) based code \cite{PhysRevB.53.R10441,0953-8984-14-11-302}. A double-$\zeta$ polarized basis set within the local density approximation (LDA) approach for the exchange-correlation functional was used. 
It is well known that even though the LDA is able to reproduce most of the interesting physics of a system, it underestimates the size of the band gap. Indeed, spectroscopy based measurements revealed that the band gap of phosphorene ($\sim$2.2 eV) is larger than that reported here within the DFT/LDA framework  \cite{ISI:000355620000013}.
Also the largest uncertainty in predicting the accurate position of a defect state has been a problem related to the band gap as computed with semi-local DFT xc-functionals. One must assume some uncertainty when dealing with systems composed of hundreds of atoms while obtaining a coherent description of it. In this case, accuracy is given up in favor of performance by choosing a functional that makes the study achievable while reliable.

2D and 1D monolayered phosphorene sheets were modeled within a supercell large enough to avoid interactions between neighboring cells. Atomic positions were relaxed with a force tolerance of 0.02 eV/\AA. The integration over the Brillouin zone was performed using a Monkhorst sampling of 4$\times$4$\times$1 k-points for 3.6nm$\times$3.2nm supercells containing a defect. 1x1x4 k-points were used for computing the Hamiltonians of 14-primitive zigzag unit cell long defective ribbons. The radial extension of the orbitals had a finite range with a kinetic energy cutoff of 50 meV. The numerical integrals were computed on a real space grid with an equivalent cutoff of 300 Ry. 
 
\subsection{Electronic-transport approach}
To determine the electronic-transport properties of modified phosphorene ribbons we used the Landauer-B\"uttiker formulation of the conductance, which is particularly convenient to analyze the electron transport along a 1D device channel in between two semi-infinite leads. The scattering region where charge carriers can be backscattered during their propagation is a defective phase-coherent multi-mode channel in between two semi-infinite leads (pristine ribbons), which are in thermodynamical equilibrium with infinitely larger electron reservoirs. The computational strategy  \cite{Biel, swnt} used for the description of the electronic-transport properties is as follows. A set of first-principles calculations are performed first to obtain the {\it ab initio} Hamiltonians (H) and overlap (S) matrices associated with chemically modified or defective ribbon segments. The atomic-like basis set utilized by the SIESTA code allows us to obtain a description of the impurity potential with relatively small and manageable sparse Hamiltonian matrices. The modified ribbon unit cells are long enough so that the ribbon-extremes are well converged to the clean system. Thus functionalized and clean segments of phosphorene ribbons can be assembled to construct long systems with perfect contact areas between the building blocks. Pieces of both modified and pristine segments are connected in a random manner to mimic translational disorder. Within a O(N) scheme with respect to the ribbon length, standard real-space renormalization techniques to calculate the Green functions associated with the sparse Hamiltonians and overlap matrices are further employed to include recursively the contribution of the ribbon segments. Thus quantum transport property analysis of micrometer-long ribbons can be attained within the accuracy of first-principles calculations. To evaluate the conductance $G$ of the system we adopt the standard Green function formalism which, at quasi-equilibrium conditions, is $G(E)={G_0}\sum_{n}T_n(E)$, where $G_0=2e^{2}/h$ is the quantum of conductance. The transmission coefficients $T_n(E)$ are calculated by evaluating the retarded (advanced) Green functions of the system:
\begin{equation}
\mathcal{G}^{\pm}(E)=\{E S-H-\Sigma^{\pm}_{L}(E)-\Sigma^{\pm}_{R}(E)\}^{-1}
\end{equation}
where $\Sigma^{\pm}_{L(R)}(E)$ are the self-energies which describe the coupling of the channel to the left (right) leads. These quantities are related to the transmission factor by the formula  \cite{0022-3719-4-8-018}:
\begin{equation}
T(E)=tr\{\Gamma_{L}(E) \mathcal{G}^{+}(E) \Gamma_{R}(E) \mathcal{G}^{-}(E)\}
\end{equation}
with 
$\Gamma_{L(R)}(E)=i\{\Sigma^{+}_{L(R)}(E)-\Sigma^{-}_{L(R)}(E)\}$. $tr$ stands for the trace of the corresponding operator. 

The transmission $T_n(E)$ for a given conducting channel $n$ gives the probability of an electron to be transmitted at energy $E$ from the source to the drain electrode. For a pristine ribbon, $T_n(E)$ acquires integer values corresponding to the total number of open propagating modes at the energy $E$. For an isolated ribbon, $T_n(E)=1$ near the Fermi level in the electron band (first plateau) and 3 for the second plateaus (a tiny plateau of 2 transmitting channels is also present in the hole band). The conductance of a pristine phosphorene nanoribbons in the ballistic conduction regime is given as the sum of all conducting channels at a given energy. 

A quantitative estimation of the final transport gaps necessitates a self-consistent evaluation of the charge flow to account for the accumulation of charges at the ribbon, which in turn screens the impurity potential. Additional analysis based on this effect might provide further information on the gaps, as well as the effect on the transport characteristics of impurities when they become charged  \cite{PhysRevB.82.115318}.

\section{Results and discussion}
Monolayered phosphorene flakes can be exfoliated micromechanically on a SiO$_2$ substrate used as the back gate of field effect transistors. 
Structural defects as a result of the damage induced during the sample preparation, functional groups in the environment that attach to the surface, or purposely substitutional doping are the major source of material surface perturbation. 
We wil start by presenting a model to explain the particular orbital hybridization of the P atoms. Next, one by one different scenarios of phosphorene chemical modification will be considered, paying attention to structural, electronic and magnetic changes introduced upon H atom attachment, row-II and -III elements doping, and di-vacant site interstitial atoms. The consequences of each type of modification on the quantum transport properties of long phosphorene ribbons will be then analyzed. 

\subsection{Pristine phosphorene}

\begin{figure}[htp]
 \centering
	\includegraphics[width=0.5 \textwidth]{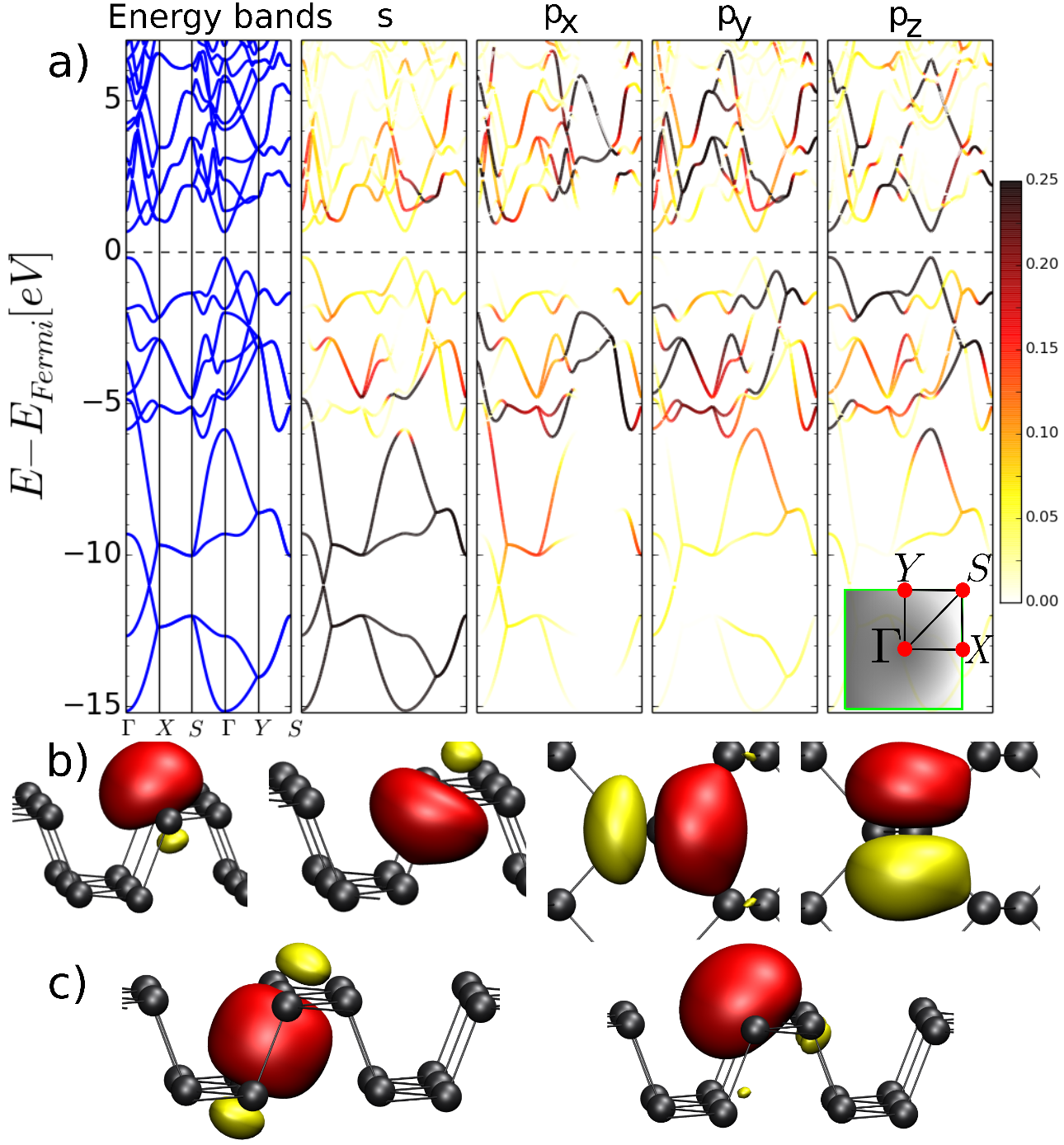}  
	\caption{ a) Energy band diagram of phosphorene along the path joining the high-symmetry points: $\Gamma \rightarrow X \rightarrow S \rightarrow \Gamma \rightarrow Y \rightarrow  S$. The color-weighted representation of the contribution of each s and p orbital shows a strong mixing of the states at all energies. b) isosurfaces (2e$^-$/\AA$^3$) of a $\sigma$-type and sp-like type MLWFs. c) Isosurfaces corresponding to the MLWFs projected onto both the occupied and unoccupied bands.}
	\label{fig0}
	\end{figure}

The electronic band diagram of phosphorene and the contribution of each P atomic orbital to each band are shown in Figure \ref{fig0}. Whereas the deeper bands are generated mainly by the s orbitals, the six occupied bands of higher energy exhibit a significant contribution of all orbitals. Although this orbital mixing suggests a sp$^3$ type of hybridization for all the atoms a detailed analysis of the lattice geometry leads to a different conclusion. 

The corrugated geometry of phosphorene exhibits a vertical separation between P atoms of $\sim$2.1 \AA. The bonding length for P atoms in an in-plane zigzagged line is $\sim$2.24 \AA, whereas the bonding length between atom chains at different planes is $\sim$2.26 \AA. Three out of five of the P valence electrons participate in the formation of interatomic bonds. The in-plane P-P-P angle of 95.8$^\circ$ suggests that the two p-p-$\sigma$ bonds between neighboring in-plane P atoms are constructed basically from non-hybridized p orbitals, similarly to the phosphine molecule (PH$_3$), where the P-H angle of the p-s-$\sigma$ bond is 93.5$^\circ$. On the contrary, a bond angle of 103.5$^\circ$ with the P atom in the second zigzagged chain points out to a predominantly sp-type orbital hybridization. A non-negligible component of the in-plane p$_x$ and p$_y$ orbitals allows the sp hybrid orbital to bend away from 90$^\circ$. The remaining two electrons reside as a lone pair in the second sp orbital, which yields the strong out-of-plane dipolar moment of phosphorene and explains the high hydrophilic character of the surface \cite{2053-1583-2-1-011002}. 

To fully characterize this model of P orbital hybridization we resort to the the maximally localized Wannier functions (MLWFs) \cite{PhysRevB.56.12847}, obtained using the Wannier90 code \cite{Mostofi2008685} from the first-principles ground state obtained with Quantum-ESPRESSO \cite{Espresso}. 
By projecting onto s and p orbitals and minimizing the MLWF spread, the band structure obtained using the first-principles approach compared to the Wannier interpolation was in excellent agreement. 

Figure\ref{fig0}b shows the real-space representation of the four types of Wannier functions constructed by considering the ten Bloch states up to the Fermi-level plus six unoccupied bands. As a result, the two equivalent in-plane p-type orbitals and the two sp-type orbitals centered on each atom predicted above are obtained.

To obtain further insight into the local chemistry of the phosphorene monolayer, MLWFs are employed to form a realization of the interatomic bonds and electron lone pairs. Projecting only onto the ten fully occupied bands the resulting MLWFs are a set of six symmetry-related bond-centered $\sigma$-orbitals and four atom-centered s orbitals (Figure\ref{fig0}c). A projected density of states onto each of the four types of Wannier functions reveals that all contribute similarly to the occupied bands at all energies due to the strong hybridization between P orbitals. Note that this particular representation applies only to the valence bands, and it represents the strongly localized nature of the $\sigma$-bonds joining the P atoms and the out-of-plane orbital containing the lone pair, in a picture that resembles that of the phosphine molecule. 

\subsection{Mono- and di-hydrogenated phosphorene}

\begin{figure}[htp]
 \centering
 \includegraphics[width=0.5 \textwidth]{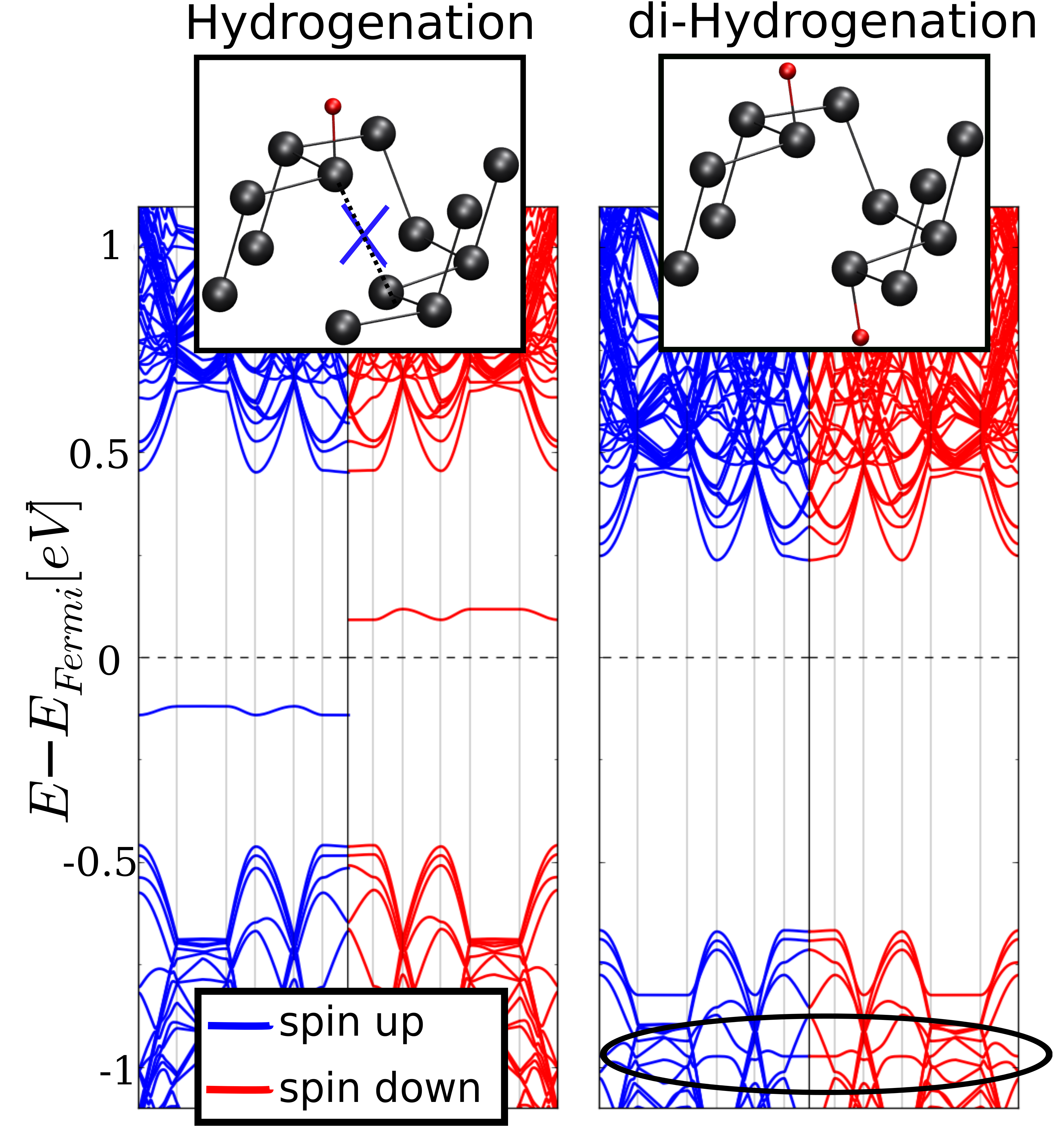}  
 \caption{Left panel: spin-resolved band diagram of a 3.8 $\times$ 3.1 \AA\ mono-hydrogenated phosphorene supercell. The inset shows the broken bond upon H attachment. Right panel: similarly for a two-fold hydrogenated phosphorene supercell where the two H are attached to consecutive P atoms at opposite zigzagged chains. The two flat states in resonance with the valence bands are pointed out by an oval. The bands are plotted along the path joining the high-symmetry points: $\Gamma \rightarrow X \rightarrow S \rightarrow \Gamma \rightarrow Y \rightarrow S/Y\rightarrow X$, pointed out by the vertical lines.}
 \label{fig1}
\end{figure}

The PH$_3$ molecule orbital model is locally recovered in phosphorene when a monovalent H atom attaches a P atom. To form the new P-H bond 1.5 \AA\ long, the P atom is pulled slightly out of the surface by 0.14 \AA\ increasing the in-plane P-P bond angle to 99.1$^\circ$. The H-P-P angle of 90.8$^\circ$ is close to the angle 93.5$^\circ$ of the P-H phosphine bond. It is reasonable to assert that the atomic orbitals of the modified P atom are not hybridized. The P atom in the chain beneath feels the repulsion of the lone pair electrons above and is displaced outwards from the plane by 0.25 \AA. The inter-chain bond is broken and the atom is left with only two neighbors (see the inset of the leftmost panel of Figure \ref{fig1}). 

In the hypothetical case that the hybridization of the P orbitals in phosphorene was principally sp$^3$, the H would attach to the P atom forming a large angle with the surface in order to maintain the tetragonal geometry of the diamond-like structure. By forcing the H atom to adopt this orientation a fully relaxed DFT calculation yields a H-P-P bond of 102.5$^\circ$ and a P-P-P bond of 104.0$^\circ$, very close to the characteristic 109.5$^\circ$ of the sp$^3$ hybridization. The P atom to which the H atom is attached is displaced inwards from the structure by 0.3 \AA, and it pulls the P atom in the chain below out of the plane by 0.2 \AA undergoing a sp$^3$ rehybridization according to the new P-P angles in the range from 103$^\circ$ to 110$^\circ$. This strong deformation of the network increases the configuration energy by 0.1 eV with respect to the previous one. In a band diagram representation the flat states are at high energies very close to the conduction band.  

The change from p-p-$\sigma$ to p-s-$\sigma$ bonding upon attachment of the H atom is accompanied by the formation of a local magnetic moment of 1 $\mu_B$.
This is the result of an unpaired electron that remains in the vicinity of the new $\sigma$ bond and has a very low spatial extension, as inferred from the band diagram of Figure \ref{fig1} by a filled in-gap non-dispersive state. The unsatisfied orbital (a hole) is reflected as an antibonding state. The electronic bands are plotted along high-symmetry lines of the Brillouin zone of a 3.8 $\times$ 3.1 \AA\ supercell. 

Similarly to highly reactive molecules containing an unpaired electron in the outer orbital, this radical is unstable and tends to react readily with other species. A second H atom forms a stable bond with the P atom in the opposite zigzagged chain to which the first modified P atom was attached (see the inset of the rightmost panel of Figure \ref{fig1}). After the new p-s-$\sigma$ bond is created, the distance between both P atoms increases to 2.9 \AA. As the band diagram of Figure \ref{fig1} shows, the attachment of a second H atom quenches the radical and lowers the energy of the non-dispersive $\sigma$ states which are in resonance with the valence bands at -0.97 eV (denoted in the Figure by an oval).

A second H atom attached to a neighboring P atom in the same chain removes the unpaired electron as well. This creates a doubly occupied in-gap state, 0.25 eV above the valence band, at the expense of inducing a lattice distortion that increases the configuration energy by 1.02 eV with respect to the latter configuration. The determination of the correct distribution of defect states is important for the transport properties since they are responsible for antiresonant scattering mechanisms.

\subsection{Doping} 

\begin{table*}
\caption{Distances from doping atom to the two equally distant P atoms, d(P$_1$), and to the third neighboring atom, d(P$_2$). $\alpha_1$ ($\alpha_2$) indicates the angle between the doping atom and the P$_1$ (P$_2$) atom.}
\label{table1}
\begin{tabular}{llllllllllll}
\hline
& \multicolumn{1}{c}{B}   & \multicolumn{1}{c}{C}  & \multicolumn{1}{c}{N}  & \multicolumn{1}{c}{O}  & \multicolumn{1}{c}{F} & \multicolumn{1}{c}{Al}  & \multicolumn{1}{c}{Si}  & \multicolumn{1}{c}{P}& \multicolumn{1}{c}{S}  & \multicolumn{1}{c}{Cl}  \\ \hline
d(P$_1$)[\AA]	& 1.9& 1.8& 1.8& 1.8& 2.9& 2.3& 2.9& 2.25& 2.2& 2.7\\ 
d(P$_2$)[\AA]	& 1.8& 1.8& 1.8& 2.8& 1.7& 2.3& 2.9& 2.25& 2.8& 2.3\\ 
$\alpha_1$[$^\circ$] & 123.3& 103.4& 119.3& 113.6& 98.7& 115.3& 107.0& 103.6& 105.1& 131.5\\ 
$\alpha_2$[$^\circ$] & 109.1&  95.5& 103.2& 116.9& 55.5&  98.2&  96.3&  95.5& 102.2&  83.4\\ 
\hline
\end{tabular}
\end{table*}

\begin{figure*}[htp]
 \centering
 \includegraphics[width=0.99 \textwidth]{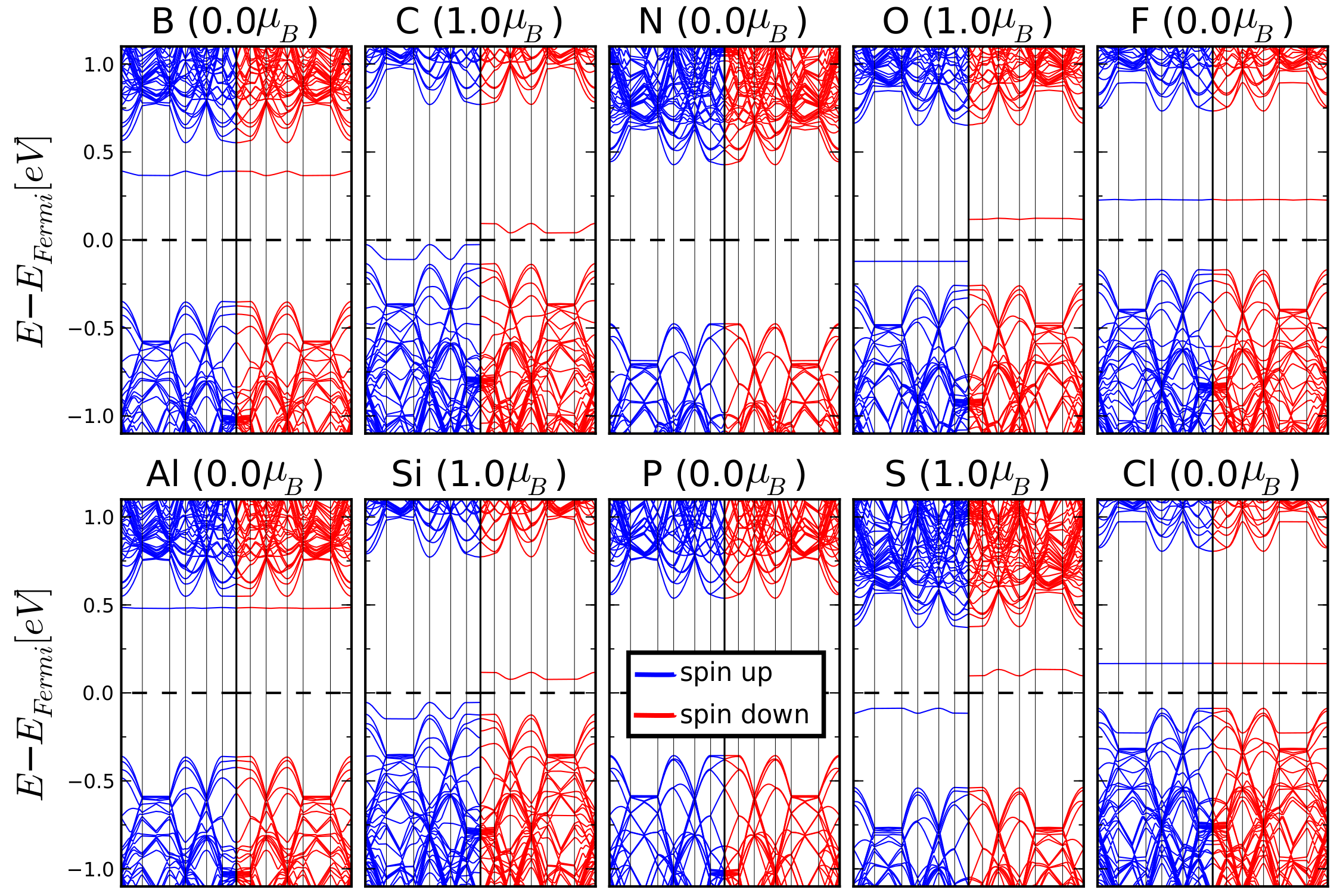}  
 \caption{DFT-computed spin-resolved band diagram of a 3.8 $\times$ 3.1 \AA\ phosphorene supercell with substitutional doping of B, C, N, O, and F atoms (upper panels) and with Al Si, S, Cl atoms (lower panels). The central panel in the lower row corresponds to the pristine phosphorene supercell. The bands are plotted along the path joining the 
 high-symmetry points: $\Gamma \rightarrow X \rightarrow S \rightarrow \Gamma \rightarrow Y \rightarrow S/Y\rightarrow X$, pointed out by the vertical lines. }
 \label{fig2}
\end{figure*}

The process of doping, by which a foreign element is introduced in a material in substitution of another atom to alter its electrical properties, is commonly used in the semi-conducting industry to add or remove charge carriers. The five valence electrons of P have traditionally been used for Si n-type doping. B, which has three valence electrons, is used for Si p-type doping. 
In monolayered materials p- and n-type doping achieved through atom substitution and surface transfer doping are the most practicable method to control free carrier concentrations . Both B and N have been demonstrated to be successful pathways to dope graphene, adding new functionalities to the metallic layer, such as the creation of transport gaps \cite{PhysRevLett.102.096803,Roche2012960}. 
Hydrogenation, oxidation, or fluorination \cite{PhysRevB.81.205435,PhysRevLett.103.056404} are able to turn graphene from a weakly disordered semimetal to a wide-band-gap insulator depending on the density of the impurity atoms. 

Due to the demonstrated effectiveness with which neighboring atoms in the Periodic Table combine, second- and third-row elements are good candidates to tune the electronic and transport properties of phosphorene. This technique can be used in conjunction with material thickness to tackle its effective gap. 
Here, doped phosphorene was constructed by locating a single foreign atom in a large phosphorene supercell so that the impurity is truly isolated from the neighboring virtual images. This prevent the impurity states from overlaping and creating an impurity band, and the material remains a non-degerate semiconductor. As for H attachment, the minimal doped system was constructed by repeating 11 $\times$ 7 times a phosphorene unit cell. A single doping atom in substitution of a P atom was observed always to be immersed into the phosphorene puckered layer, displacing the surrounding P atoms away from their original positions. 

In Table \ref{table1} complete information on the distances and angles between each doping atom and the surrounding P atoms is reported. Group III and IV foreign atoms are practically equidistant to the three neighboring P atoms. The small radius B and C atoms shorten the original P-P bond length, whereas Al and Si substitutional atoms create longer bonds. Both O and S atoms create shorter bonds with the two in-plane atoms than with the third P atom in the opposite chain. Interestingly, a F atom sits on top of a P atom at 1.73 \AA, similarly to a H adatom, whereas Cl establishes tighter bonds with the two other P atoms.

Figure \ref{fig2} shows the DFT-based computed electronic spin-resolved band diagrams of monolayered black phosphorus modified with B, C, N, O, and F atoms for the upper panels, and with Al Si, S, Cl atoms for the lower panels. The band structure of pristine phosphorene is also included as a reference.
The densely packed lines in the band diagrams are the result of the multiple band backfolding obtained upon increasing the lateral size of the supercell up to 3.8 $\times$ 3.1 \AA. The electronic and magnetic effects are clearly visible in the band diagrams of Figure \ref{fig2}, where three different cases can be distinguished depending on the number of valence electrons of each doping atom. 

First, there is a remarkable the strong resemblance between the N-doped phosphorene band diagram and that of pristine phosphorene. The degeneracy of the electronic states of the pristine material is barely removed when the isoelectronic and smaller N atom is substituted for a P atom. The N is displaced inwards from the structure to reduce the N-P bond length by a 18\%. A similar behavior was observed for an As atom in substitution. 

A completely different effect is observed for doping atoms with an even number of valence electrons. As shown in Figure \ref{fig2}, a flat state below the Fermi level has its corresponding anti-bonding state in the opposite spin panel, and always within the electronic band gap. Similarly to the H atom attachment procedure analyzed above, as a consequence of the unbalanced number of electrons a non-dispersive radical centered in the vicinity of the impurity is created and accounts for the net magnetic moment of 1$\mu_B$ observed in all group IV and VI atoms. Both C and Si remove most of the band degeneracy of the pristine material electronic states, introducing strong distortions in the bands that are more remarkable in the spin panel containing the defect filled band. 

A third type of doping considers the substitution with atoms containing an odd number of electrons in the outer shell. Both B and Al employ their three valence electrons on creating chemical bonds with the three neighboring P atoms. An in-gap dispersionless state for each spin configuration is located at high energies in resonance with the conduction band states. Similarly, the halogen atoms create two flat empty bands but at lower energies. 
The impurities are paramagnetic as demonstrated in the band diagrams by all the electronic states being symmetric across the two spin panels.   
Therefore, it can be concluded that there exits a clear interplay between the outer shell configuration and the electronic and magnetic properties of the doped system as one moves across the second and thrid rows of the periodic table. 

It is worth noting that the magnetism arising in defective phosphorene is related to the p or hybridized orbitals. 
As in graphene and related C-based hyperconjugated materials, the magnetic moments are completely derived from unpaired $\pi$-electrons. Unlike the d- and f-electron systems in which the positive exchange interaction is derived from the orthogonality and the space-sharing of the atomic orbitals, in p orbital based systems the spin lives around the doping or adsorbed atom in $\pi$-type orbitals whose extension depends on the nature of the impurity.

\subsection{Interstitial doping atoms}
Imaging techniques based on high-resolution transmission electron microscopy have been widely employed for the characterization of multiple types of monolayered compounds, ranging from graphene to h-BN or dichalcogenides \cite{PhysRevLett.109.035503}. However, due to the interaction of the high energy electrons with the atomic species as they are transmitted through the ultrathin sample, the production of defects such as mono- and multi-vacancies is an unavoidable consequence. Also, the mechanical exfoliation or interaction of the samples with external agents during their preparation can be the origin of vacant sites which, in turn, may work in the material's favor. 
Indeed, the removal of P atoms leads to structural deformations and unsaturated bonds, which enhance the chemical reactivity of the defect and provides an anchor site for external doping atoms or functional molecules. 
Topological defects may enhance the sensitivity of P-based chemical sensors towards molecule deposition due to an efficient adsorption and an enhanced charge-transfer process. The formation energy for external atoms adsorbed on defective phosphorene has been reported to have increased stability compared to pristine phosphorene \cite{doi:10.1021/jp5110938}. This type of defects must also be considered as a fundamental source of scattering centers, and it is responsible for the lower mobility values observed in black phosphorus, as will be demonstrated in the next section.

\begin{figure*}[htp]
 \centering
 \includegraphics[width=0.99 \textwidth]{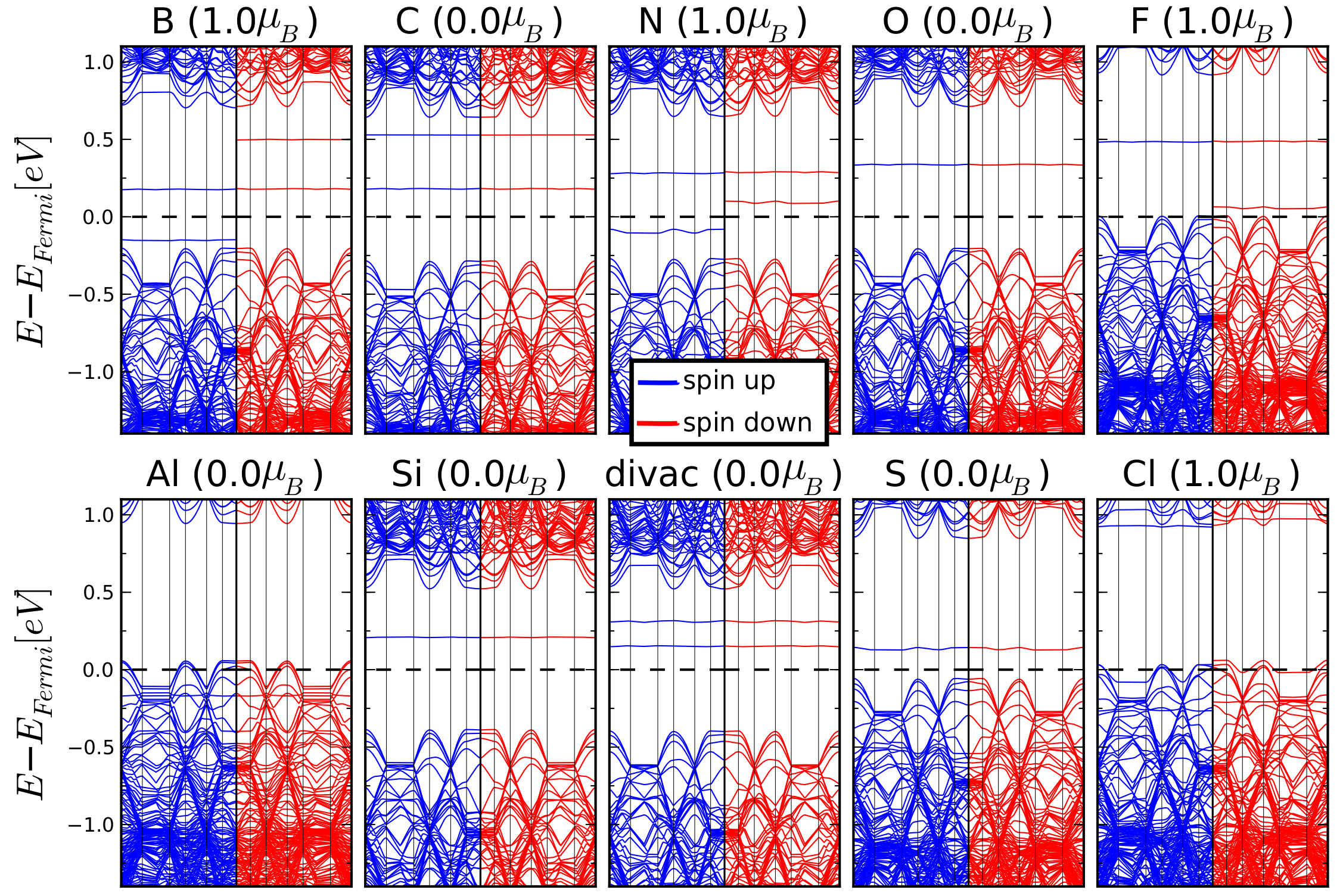}  
 \caption{DFT-computed spin-resolved band diagram of a 3.8 $\times$ 3.1 \AA\ phosphorene supercell with interstitial doping of B, C, N, O, and F atoms (upper panels) and with Al Si, S, and Cl atoms (lower panels). The band diagram of the phosphorene divacancy is plotted in the central panel in the lower row. The bands are plotted along the path joining the 
 high-symmetry points: $\Gamma \rightarrow X \rightarrow S \rightarrow \Gamma \rightarrow Y \rightarrow S/Y\rightarrow X$, pointed out by the vertical lines. }
 \label{fig3}
\end{figure*}

Due to the corrugated geometry of phosphorene on P zigzag chains linked in a staggered lattice, the number of di-vacancy geometries is considerably higher than in other monolayered materials \cite{0957-4484-26-6-065705}. The most stable phosphorene di-vacancies as described in Ref  \cite{0957-4484-26-6-065705} will be considered here as the anchor site where a doping atom attaches to the P network. The removal of a pair of P bonded atoms belonging to two different zigzagged chains leads to the formation of an irregular octahedron surrounded by six hexagons and two pentagons. Every P atom conserves the same coordination of the original structure although four atoms enlarge the bonding angle with their first neighbors. This strong distortion is used by the doping atoms to attach the P network with different geometries depending on the atom electronegativity and radii. One of the pentagons transforms into a hexagon upon attachment of B, C, or N atoms to the longest P-P bond of the vacant site. On the contrary, row-III atoms attach to four P atoms simultaneously creating two new hexagons and two pentagons. 

A single dopant located as an interstitial atom at a di-vacancy affects the electronic and magnetic properties of the phosphorene layer in a different fashion than in substitution or chemisorption. The electronic band diagrams in Figure \ref{fig3} summarize the effect of row-II and -III atoms. The electronic states of a fully relaxed di-vacancy are shown for comparison. 
Extracting a P dimer from the pristine structure of phosphorene enhances the hybridization of the electronic states, removes the band degeneracy, and causes the appearance of two in-gap empty states for each spin channel. One of the main consequences of attaching foreign atoms to the defect site is the shift of these states either closer to the conduction bands or below the Fermi level. Phosphorene becomes magnetic upon interstitial attachment of a trivalent B atom as a consequence of the filling of one of these states, whereas the corresponding antibonding state is located at higher energy. A N atom, isovalent to P, leads to a similar behavior of the defect states, as does a F atom which locates the defect filled state in resonance with the valence bands at -0.25 eV. In the three types of doping a net magnetic moment of 1$\mu_B$ is induced.

A different behavior is observed for the atoms of the III-row. Both Al and Cl atoms sitting in the interstice lead to p-type doping, with Al locating the defect states at -0.2 eV and -0.4 eV in the valence bands and Cl at about -0.25 eV and 0.9 eV. An unbalanced repartition of the spin-resolved charge density yields a net magnetic moment of 1$\mu_B$ for the latter.
The atoms with an even number of valence electrons induce no magnetic moment when sitting in the interstice, and only C doping allows the original two sets of flat states per spin channel to remain in the band gap.

\subsection{Quantum transport properties of hydrogenated phosphorene ribbons}

A characteristic feature of black phosphorus is that an electron-hole current asymmetry can be induced by modifying the number of layers \cite{doi:10.1021/nl5025535}. Alternatively, other methods based on the chemical and structural modification of the phosphorus network could be employed to purposely tune its electronic band gap or explain its experimentally measured low mobility.
A disordered distribution of impurities creates energy levels made of low or non-propagating states as a result of the fluctuations of the material local potential with the disordered distribution of impurities. Free charge carriers in these states cannot propagate, and this region of energy states is identified as constituting a transport gap. 

The scope of this paper is to show the impact of the chemical and structural modifications analyzed above on the quantum transport properties of monolayered black phosphorus. In particular, we will focus on phosphorene flakes of reduced lateral width. This allows the antiresonant states to broaden the energy range where scattering occurs, providing a clearer insight into the effect of a single impurity on the transmission probability of a charge-carrier along the ribbon . The first-principles calculations performed in two dimensional phosphine slabs were further reproduced in one-dimensional ribbons. 

The minimal atomic model of a modified phosphorene ribbon was constructed by repeating a 12 P-dimer zigzag unit cells 14 times along one axis such that the geometric and energetic perturbations caused by the defect vanish at the edges of the supercell. To avoid any interference from the dangling bonds of P edge atoms, the edges were passivated with a H atom, which in turns lowers the configuration energy with respect to the bare ribbon \cite{PhysRevB.90.085424}.

\begin{figure}[htp]
 \centering
 \includegraphics[width=0.45 \textwidth]{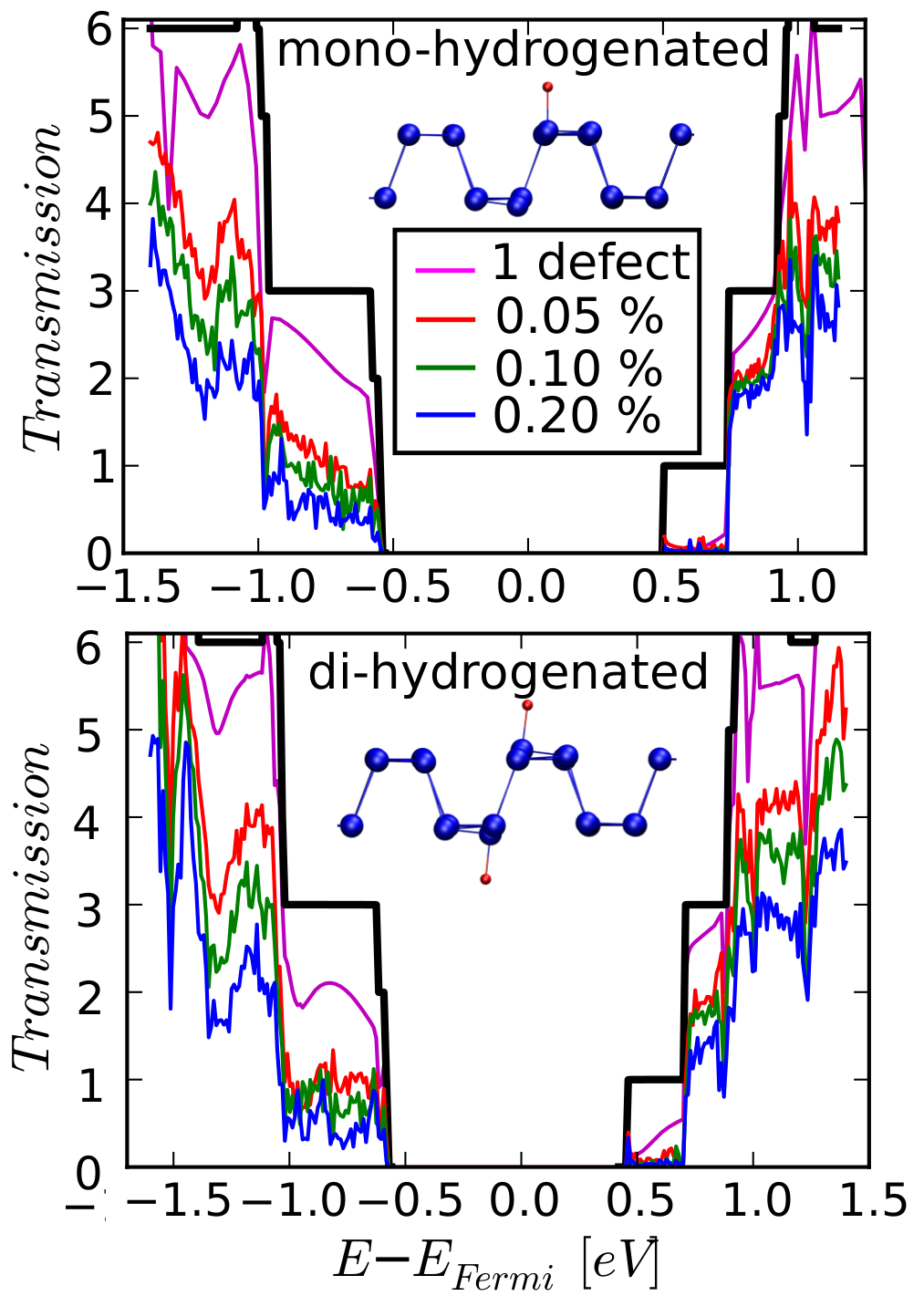}  
 \caption{Transmission profiles for phosphorene ribbons modified with one single H atom, (left panel) and with two H atoms (right panel) attached to opposite P atoms at both sides of the surface, as shown in each panel inset. For an increasing concentration of defects in a 1 $\mu$m long ribbon the transmission rapidly enters the localized regime in the vicinity of the gap, and the diffusive regime at higher absolute energies.}
 \label{fig4}
\end{figure}

Firstly, the consequences of mono- and di-hydrogenation on the charge-transport properties of phosphorene ribbons are examined. The attachment of a H atom is an example of a single-bond chemisorbed specie. The upper panel of Figure \ref{fig4} shows the transmission profiles of an infinite ribbon with an increasing number of single H atoms attached to the surface. The stepwise lines spanning an energy range of 3 eV around the Fermi level correspond to the transmission profile of a defectless infinite ribbon. The sum of the conducting modes at each energy yields the characteristic conductance plateaus of a pristine 1D network. 

A substantial transmission drop is observed (purple curve) if a single H atom is attached to the surface.
While the loss of transmission is $\sim$10	\% on average for the plateaus of higher energy, the first plateau exhibits a full suppression of the transmission which is partially recovered after the onset of the next conducting mode. This electron-hole asymmetry is more evident when the accumulated effect of an increasing number of atoms randomly distributed over 1 $\mu$m long ribbons further drops the conductance. Indeed, due to disorder effects and quantum interferences between the quasi-bound states created by the defects, the transmission drop of one single impurity is highly amplified when adding a much larger number of scattering centers. A statistical analysis was performed by averaging the computed transmission coefficients over 10 disordered configurations. The mesoscopic transport calculations were performed introducing 0.05\%, 0.10\%, and 0.20\% of defects with respect to the number of P atoms.

The lower panel of Figure \ref{fig4} shows a similar analysis for the di-hydrogenated ribbon. Despite the fact that the attachment of the second H atom additionally distorts the P network and removes the high-energy radical, an electron-hole asymmetry similar to the mono-hydrogenated ribbon is observed. One could expect that once the radical is quenched the transmission is recovered. To the contrary, the di-hydrogenation allows the system to enter the localized regime for lower defect concentrations in the energy range from -1.0 to 0.75 eV, as observed with transmission values below the limit of T=1. However, while the mono-hydrogenation saturates the transmission drop in the electron band rapidly with an increasing number of defects, the di-hydrogenation reduces the transmission more gradually with the defect concentration. This is explained by the presence of the defect states at $\sim$-1 eV in resonance with the valence bands, whereas the single H defect is located in the flat states within the gap (see Figure\ref{fig1}). 

As we will see below for doping elements, the charge carriers do not interact with the in-gap defect states due to the absence of available electrons at the energies where those states are located. Whereas in graphitic systems the conductance is highly sensitive to the presence of unpaired electrons that interfere with the propagating charge-carrier  \cite{Dubois}, phosphorene is less affected by the mono- or di-hydrogenation, which represents an advantage towards the design of an experimental device.

\subsection{Quantum transport properties of doped phosphorene ribbons}

Substitutional doping has been demonstrated successfully as a route for creating transport gaps in nanostructures  \cite{PhysRevLett.102.096803}, and it is proposed here for modulating the intrinsic gap of phosphorene. In the following we will focus on the consequences that the most representative types of doping have on the charge-transport of micrometer long phosphorene nanoribbons. 

Both n-type and p-type doping introduce allowed energy states within the band gap and very close to the energy band that corresponds to the type of doping atom.
With a valence electron less than a P atom, a quadrivalent impurity such as a Si atom behaves like an acceptor (p-type), and a depletion of the charge density occurs in the area of the doping site. As observed in Figure \ref{fig2}, the spin-polarized defect states rely closely on the valence band. 
On the contrary, a S atom has an extra electron (n-type) with respect to a P atom and behaves as a donor element, creating an excess of charge density in the S-rich areas.
Consequently, n-type doping introduces electronic localized states way above the valence band. 
A third type of doping element is considered, namely substitution with N atoms, which is isoelectronic to P but behaves effectively as an acceptor species due to its larger electronegativity. An analysis based on the Mulliken population confirms that a charge equivalent to $\sim$0.25 electrons is gained by the N at the expense of the surrounding atoms. However, as discussed above, no localized states are observed in the electronic band gap.

\begin{figure}[htp]
 \centering
 \includegraphics[width=0.5 \textwidth]{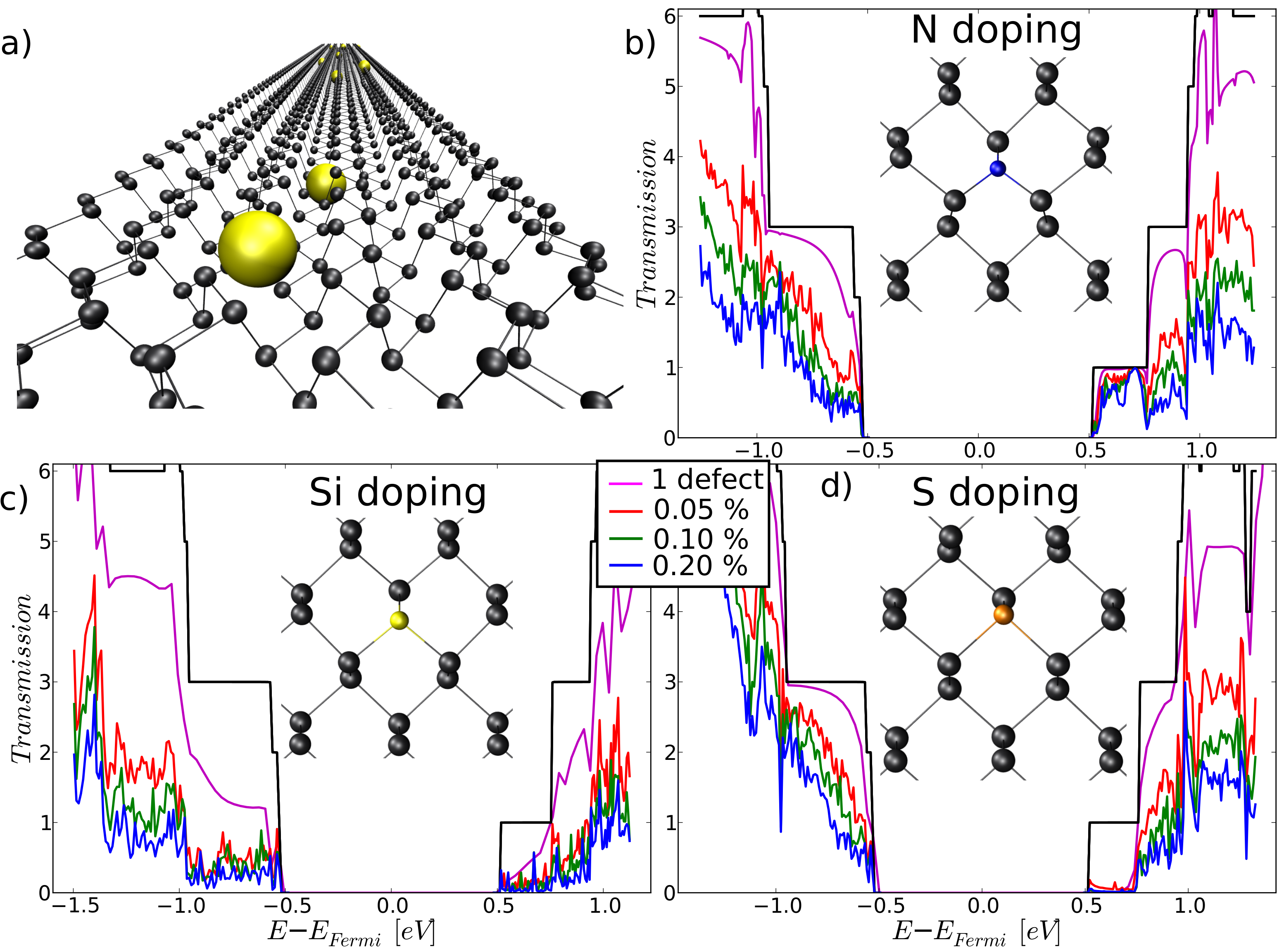}  
 \caption{Schematic representation of a long phosphorene ribbon modified with doping atoms. Transmission profiles of ribbons modified with N atoms, a), with Si atoms, b), and with S atoms, c). For Si and S doping, an increasing concentration of backscattering centers leads the transmission to enter the localized regime in the vicinity of the Fermi energy developing transport gaps.}
 \label{fig6}
\end{figure}

Figure \ref{fig6}a shows a schematic representation of a 2.6 nm wide zigzagged phosphorene ribbon with 12 P dimers across the width, and doping atoms distributed randomly along the ribbon length. 
The computed transmission for a ribbon with a single N atom is indicated in Figure \ref{fig6}b with a purple line. Although the drop in transmission is not strong, it is enhanced as more scattering centers are introduced. For a doping rate of 0.05\% the conduction in the hole band remains in the diffusive regime at all energies, and it enters the localized regime for doping rates above 0.10\%. Note the strength against backscattering at the energy range of the electron band's first plateau, which barely looses 25\% of the transmission capability for the higher doping level. 

This is in contrast to that observed in the T(E) curves for S and Si doping shown in Figure \ref{fig6}c and d respectively. The strongest transmission drop occurs in the first plateau above the Fermi energy, and it is more pronounced if possible in the case of S doping. For both n- and p-type doping a transport gap that widens the intrinsic gap of the ribbon is  developed rapidly at the electron band for S doping and at both bands for Si doping. The former effectively enlarges the ribbon gap $\sim$250 meV while the latter practically eliminates the ribbon transmission ability in an energy range of $\sim$2 eV. The insets of Figure \ref{fig6} show that the structural distortions induced by the impurities are uncorrelated with the transmission drop, since the strongest distortion is observed for a N atom (see Figure \ref{fig6} insets), which is the more robust specie against backscattering, and conversely for Si atoms. The explanation can be found in the high density of states associated with the low-dispersive states in the vicinity of the electron and hole bands, which are responsible for the enhanced backscattering efficiency of charge carriers.

In both n- and p-type doping, the fingerprint of the Fano resonances created as a result of the quantum interferences of the charge carriers with the quasi-bound states is clearly visible. Indeed, the donor potential of the S atom is a quantum well, which is reflected by a more marked transmission drop in the electron band, whereas smoother and antiresonance-free transmission curves define the conductance at the hole band. Similar effects, but on the opposite bands, are observed when the quantum waveguide is modified by the potential introduced by an acceptor Si atom.

\subsection{Quantum transport properties of phosphorene ribbons with interstitial doping atoms}

\begin{figure}[htp]
 \centering
 \includegraphics[width=0.5 \textwidth]{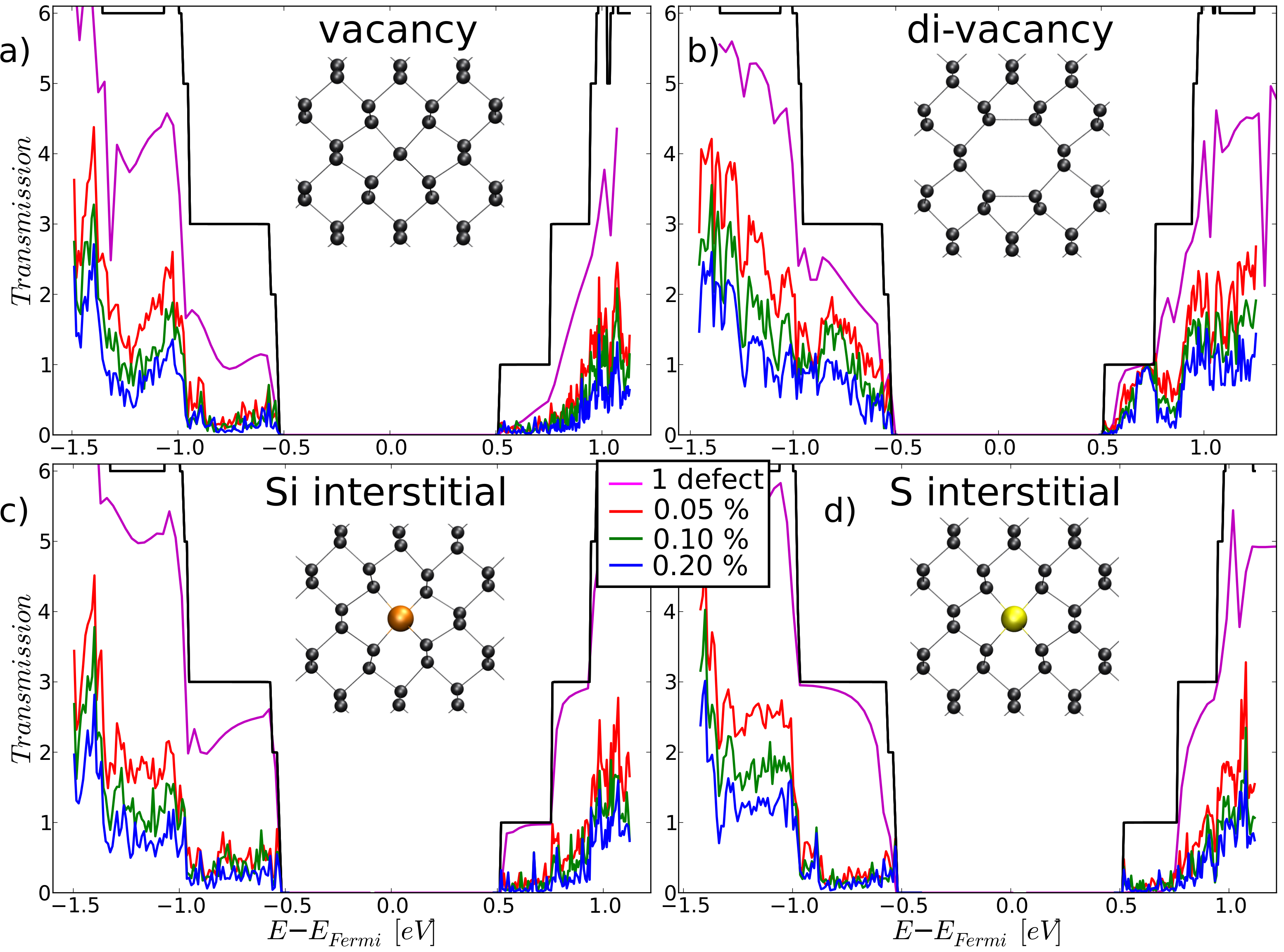}  
 \caption{Transmission profiles for phosphorene ribbons modified with a vacancy, a), with a di-vacancy, b), and with interstitial Si, c), and S, d), atoms, as shown in each panel inset. For an increasing concentration of defects the transmission drop is similar for a divancant site with interstitial atoms.}
 \label{fig7}
\end{figure}

Next we study the impact of structural defects on the transport properties of phosphorene nanoribbons. Firstly the mono-vacancy created upon removal of a P atom from the ribbon is considered. Although this type of defect as described in the Figure\ref{fig7}a-inset is considered as a transitional state towards a more stable configuration  \cite{0957-4484-26-6-065705}, it is presented here for comparison with other atoms. The transmission curve for a single defect in Figure\ref{fig7}a shows that such a perturbation of the P-network entails a drastic drop of the ribbon conductance, with a decrease of T near the Fermi energy to practically 0 in the electron band and one conducting channel in the hole band. Due to disorder effects and quantum interferences between the defect sites, the transmission drop of one single defect is highly amplified when adding a much larger number of scattering centers, leading the conductance to the localized regime even at low doping rates. 

Interestingly, removing a second P atom to create a di-vacancy partially restores the transmission. Although equal localization is reached for high doping levels, the impact of the disorder on the first transmitting modes is less strong, as Figure \ref{fig7}b shows. Similar patterns to those for the mono-vacancy are obtained when both Si and S atoms interstitially attach to the ribbon (Figure \ref{fig7}c and d). The effect of a single interstitial atom on the transmission is similar to substitutional doping, inducing the creation of transport gaps that increase effectively the band gap up to two times. Similar calculations with other elements sitting at the two-fold vacant site yielded similar results, suggesting that the transport properties of the modified ribbon are  nearly independent for this type of defect.

\subsection{Conclusions}
To conclude, a rational model supported by MLWFs has been presented to describe the hybridization of the atomic orbitals of P atoms when arranged in the black phosphorus monolayered crystalline structure. Unlike other atoms in most of the 2D family compounds, P atoms in phosphorene tend to minimize the degree of orbital hybridization, which explains the predominant angle of $\sim$90$^\circ$ both in pristine (zigzagged chains) and monovalent atom attachment (hydrogenation). First-principles electronic structure and transport calculations have been conducted on phosphorene to study the effect on the quantum transport properties of a large set of defects, ranging from surface hydrogenation to substitutional doping and interstitial atoms. Hydrogenation introduces a net magnetic moment that is quenched upon removal of the resulting radical with further attachment of a second H atom on the opposite side of the layer. Notwithstanding, and due to the new localized states that the di-hydrogenation introduces, the charge-transport ability of phosphorene is seriously affected for doping rates as high as 0.2\%. Transport gaps at both the conduction and the electron band are developed enlarging the electronic band gap of the pristine material. This is a common feature in most types of chemical modification analyzed, and it is clear evidence of the development of transport gaps in phosphorene at selective energies as a result of enhanced backscattering by impurity atoms upon the creation of dangling bonds and localized energy states. Within the micrometer scale, the contribution of quantum interferences in doped ribbons leads to a swift transition into the localized regime with increasing amount of defects. The dependency of the charge-carrier motion on the p- and n-type doping results in conductance profiles with strong electron-hole asymmetries. 

Our findings provide evidence of the sensitive nature of electronic charge-transport on a phosphorene based electronic device. A combined approach based on the incorporation of foreign species along with the stacking of multiple layers, may facilitate the design of an innovative class of all-P electronic devices.
The chemical sensitivity of the transport capability of multi-layered phosphorene with external elements distributed throughout the material, either in substitution or in interlayer intercalation, remains to be explored. Indeed, the large inter-layer separation characteristic of black phosphorus might not be an impediment for a defect on one layer to perturb the electronic properties of the neighboring layers through electronic states mixing, as it is for bilayer graphene \cite{doi:10.1021/jp5071083}, which deserves further investigation. Also, phosphorus can crystallize in different phases, including blue phosphorus  \cite{PhysRevLett.112.176802}, which may coexist with black phosphorus forming an interface  \cite{PhysRevLett.113.046804}. This which may warrant additional attention with regard to charge-transport applications.

\section{Acknowledgements}

I acknowledge DOE BES Glue funding through Grant No. FWP\#70081. I acknowledge the computing resources provided on Blues high-performance computing cluster operated by the Laboratory Computing Resource Center at Argonne National Laboratory. Work at Argonne is supported by DOE-BES under Contract No. DE-AC02-06CH11357.

\end{document}